\documentclass{PoS}
\newcommand{\msbar}{\overline {\rm MS}}
\newcommand{\SF}{Schr\"oedinger functional\ }

\title{Study of the running coupling constant in 10-flavor QCD with the Schr\"oedinger functional method
}

\ShortTitle{Study of the running coupling constant in 10-flavor QCD with the SF method
}

\author{\speaker{N.~Yamada}\,$^{,a,b}$\thanks{Email: \tt norikazu.yamada@kek.jp},
M. Hayakawa$^{c}$,
K.-I. Ishikawa$^{d}$,
Y. Osaki$^{d}$,
S. Takeda$^{e}$,
S. Uno$^{c}$
\\
\llap{$^a$} KEK Theory Center, Institute of Particle and Nuclear Studies, HIgh Energy Accelerator Research Organization (KEK), Tsukuba 305-0801, Japan\\
\llap{$^b$} School of High Energy Accelerator Science, The Graduate University for Advanced Studies (Sokendai), Tsukuba 305-0801, Japan\\
\llap{$^c$} Department of Physics, Nagoya University, Nagoya 464-8602, Japan\\
\llap{$^d$} Department of Physics, Hiroshima University, Higashi-Hiroshima 739-8526, Japan\\
\llap{$^e$} Physics Department, Columbia University, New York, NY 10027, USA
}

\abstract{
 \vspace*{-143mm}
\begin{flushright}
  \normalsize
 KEK-CP-225,\
 HUPD-0906
 \end{flushright}
 \vspace*{133mm}\
The electroweak gauge symmetry is allowed to be spontaneously broken by the strongly interacting vector-like gauge dynamics.
When the gauge coupling of a theory runs slowly in a wide range of energy scale, the theory is extremely interesting.
This may open up the possibility that the origin of all masses may be traced back to the gauge theory.
We use the \SF method to determine the scale dependence of the gauge coupling of 10-flavor QCD.
Preliminary results are reported.
}

\FullConference{The XXVII International Symposium on Lattice Field Theory - LAT2009\\
		 July 26-31 2009\\
		 Peking University, Beijing, China}

\begin{document}

\section{Introduction}

The main goal of Large Hadron Collider (LHC) is to confirm the Higgs mechanism and to find particle contents and the physics low above the electroweak scale.
So far many new physics models beyond the standard model have been proposed.
Among them, Technicolor (TC)~\cite{Weinberg:1975gm} is one of the most attractive candidates~\cite{Hill:2002ap} as it does not require elementary scalar particles which cause, so-called, the fine-tuning problem.
This model is basically a QCD-like, strongly interacting vector-like gauge theory.
Therefore, lattice gauge theory provides the best way to study this class of model\cite{Fleming:2008gy}, and the predictions can be as precise as those for QCD, in principle.

The simple, QCD-like TC model, {\it i.e.} an SU(3) gauge theory with two or three flavors of techniquarks, has been already ruled out by, for instance, the S-parmeter\cite{Peskin:1990zt} and the FCNC constraints.
However, it has been argued that, if the gauge coupling runs very slowly (``walks'') in a wide range of energy scale before spontaneous chiral symmetry breaking occurs, at least, the FCNC problem may disappear\cite{Holdom:1981rm}.
Such TC models are called walking technicolor (WTC) and several explicit candidates are discussed in semi-quantitative manner in~\cite{Dietrich:2006cm}.
Since the dynamics in WTC might be completely different from that in QCD and hence the use of the naive scaling in $N_{c}$ or $N_{f}$ to estimate various quantities may not work, the $S$-parameter must be evaluated from the first principles~\cite{Shintani:2008qe}.
Although really important quantity is the anomalous dimension of $\bar\psi \psi$ operator, to find theories showing the walking behavior is a good starting point.
Recently many groups started quantitative studies using lattice technique to answer the question what gauge theory shows walking behavior.
In~\cite{Appelquist:2007hu}, the running couplings of 8- and 12-flavor QCD are studied on the lattice using the \SF (SF) scheme\cite{Luscher:1992an}.
Their conclusion is that while 8-flavor QCD does not show walking behavior 12-flavor QCD reaches an infrared fixed point (IRFP) at $g_{\rm IR}^{2}\sim 5$.
In spite of the scheme-dependence of running and its value of IRFP, the speculation inferred from Schwinger-Dyson equation~\cite{Appelquist:1988yc} suggests that $g_{\rm IR}^{2}\sim 5$ is too small to trigger spontaneous chiral symmetry breaking.
Although 12-flavor QCD is still an attractive candidate and is open to debate~\cite{Hasenfratz:2009ea}, we explore other $N_{f}$.
In the following, we report the preliminary results on the study of the running coupling in 10-flavor QCD.
Since the lattice conference, statistics is increased by much.
The following analysis is based on the increased statistics.

\section{Perturbative analysis}

Before going into the simulation details, let us discuss some results from perturbative analysis.
In this work, we adopt the $\beta$ function defined by
\begin{eqnarray}
     \beta(g^2(L))
 &=& L\,\frac{\partial\,g^2(L)}{\partial L}
  = b_1\,g^4(L)+b_2\,g^6(L)+b_3\,g^8(L)+b_4\,g^{10}(L)+\cdots,
\end{eqnarray}
where $L$ denotes a length scale.
The first two coefficients are scheme independent, and given by
\begin{eqnarray}
     b_1
 = \frac{2}{(4\pi)^2}\left[ 11 - \frac{2}{3}N_f\right],
&\ \ \ &
     b_2
 = \frac{2}{(4\pi)^4} \left[\, 102 - \frac{38}{3}N_f\,\right].
\end{eqnarray}
The higher order coefficients are scheme-dependent and are known only in the limited schemes.
In this section, we analyze the perturbative running in the following four different schemes/approximations:
i) two-loop (universal),
ii) three-loop in the $\msbar$ scheme,
iii) four-loop in the $\msbar$ scheme,
iv) three-loop in the \SF scheme.
The perturbative coefficients relevant to the following analysis are
\begin{eqnarray}
     b_3^{\msbar}
 &=& \frac{2}{(4\pi)^6}
     \left[\, \frac{2857}{ 2}
            - \frac{5033}{18}N_f
            + \frac{ 325}{54}N_f^2\,
     \right],
\label{eq:beta-3MS}\\
     b_4^{\msbar}
 &=& \frac{2}{(4\pi)^8}
     \left[\, 29243.0
            - 6946.30\,N_f
            + 405.089\,N_f^2\,
            + 1.49931\,N_f^3\
     \right],\\
     b_3^{\rm SF}
 &=& b_3^{\msbar} + \frac{b_2\,c_2^{\theta}        }{2\pi}
                  - \frac{b_1\,(c_3^{\theta}-{c_2^{\theta}}^2)}{8\pi^2},
\label{eq:b3SF}
\end{eqnarray}
where the coefficients $c_2^{\theta}$ and $c_3^{\theta}$ depend on the spatial boundary condition of the SF used in calculations, {\it i.e} $\theta$.
Those for $\theta=\pi/5$ and $c_{2}^{\theta}$ for $\theta=0$ are known as
\begin{eqnarray}
     c_2^{\pi/5}
 &=& 1.25563+0.039863\times N_f,\\
     c_3^{\pi/5}
 &=& ({c_2^{\pi/5}})^2 + 1.197(10) + 0.140(6)\times N_f - 0.0330(2)\times N_f^2,\\
     c_2^{0}
 &=& 1.25563+0.022504\times N_f,
\end{eqnarray}
but $c_3^{0}$ is not.
Therefore, the case iv) is studied with $\theta$=$\pi/5$.
It should be noted that in our numerical simulation $\theta$=0 and thus, rigorously speaking, the example iv) is not applied to our numerical result.

%%%%%%%%%%%%%%%%%%%%%%%
\begin{figure}[tb]
\centering
\begin{tabular}{cc}
\includegraphics*[width=0.5 \textwidth,clip=true]
{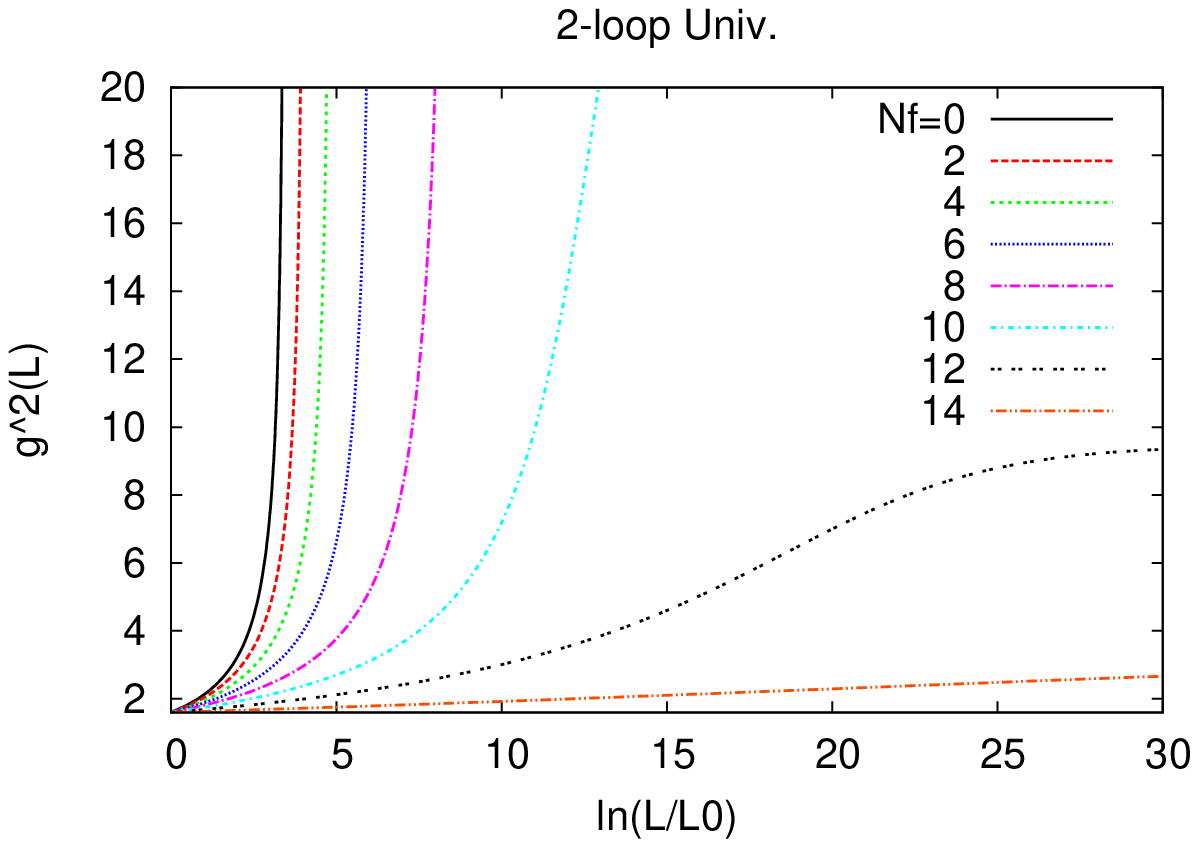} &
\includegraphics*[width=0.5 \textwidth,clip=true]
{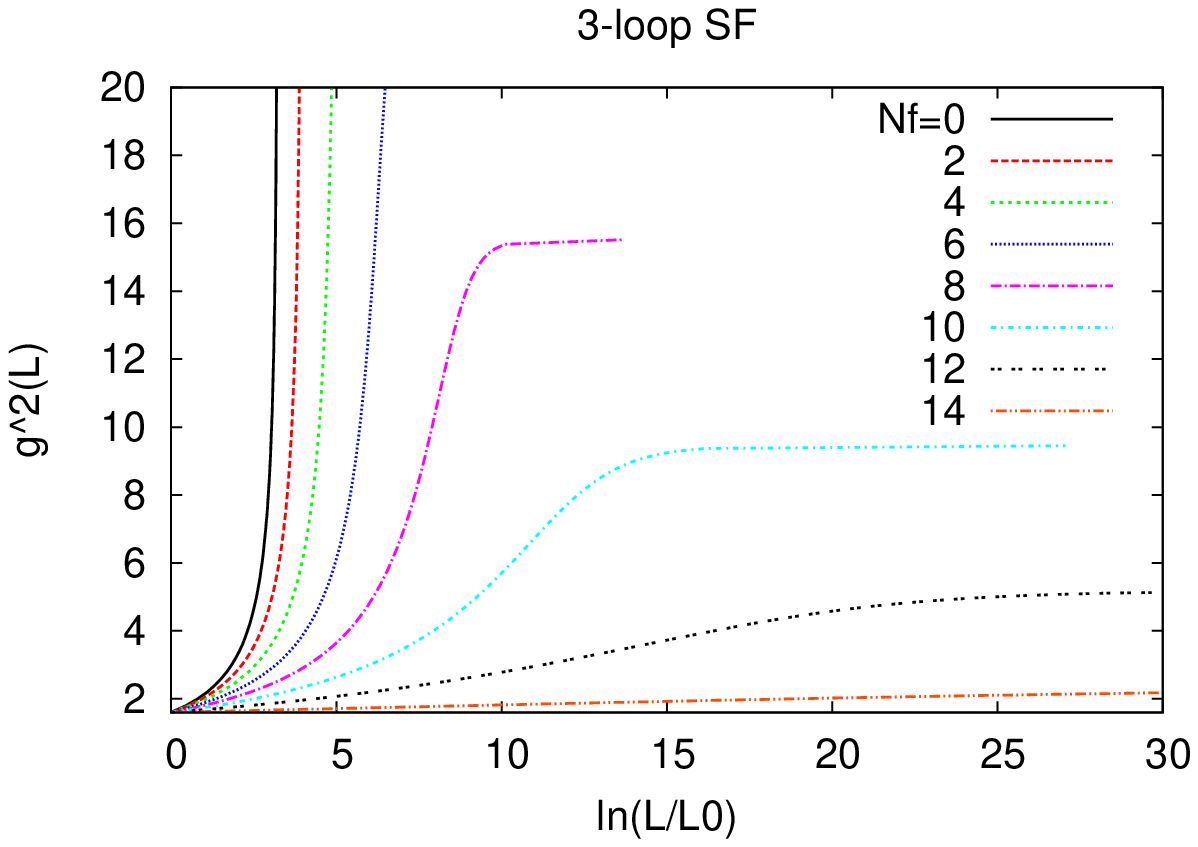} \\
\includegraphics*[width=0.5 \textwidth,clip=true]
{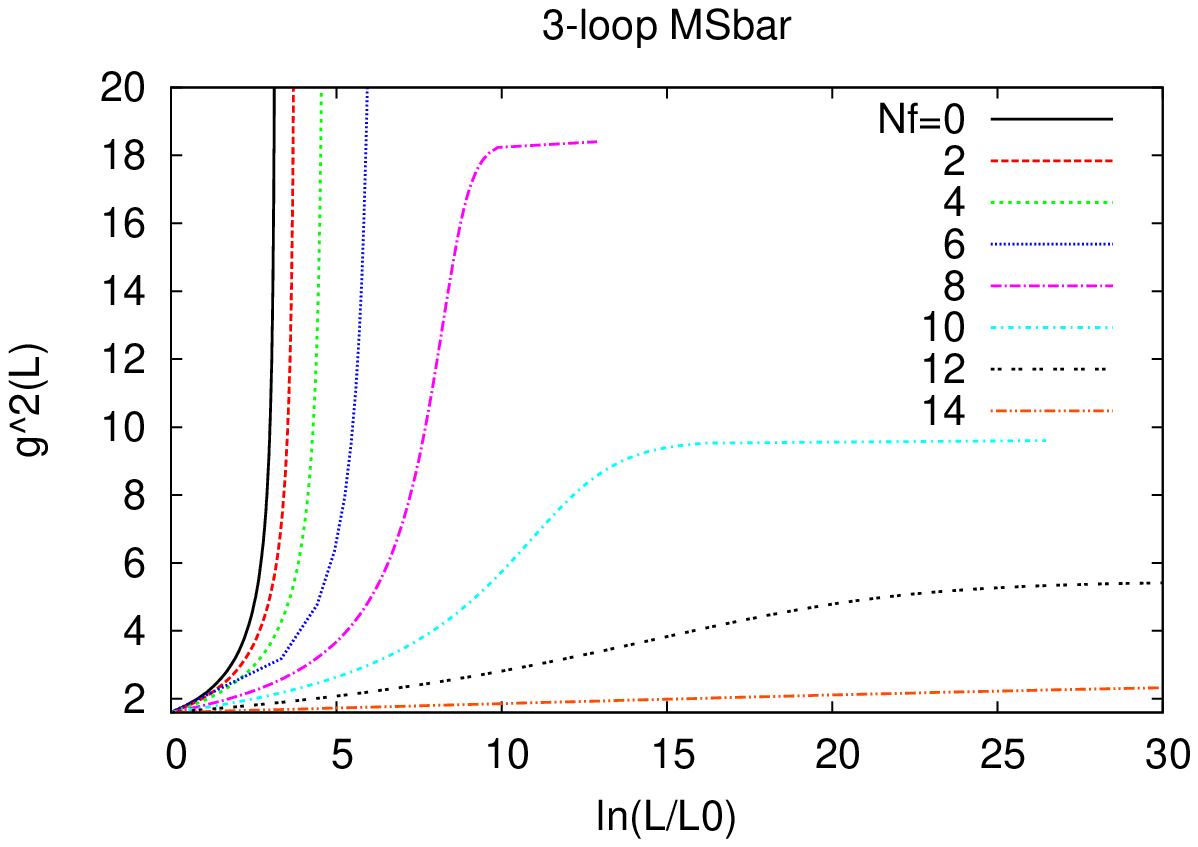} &
\includegraphics*[width=0.5 \textwidth,clip=true]
{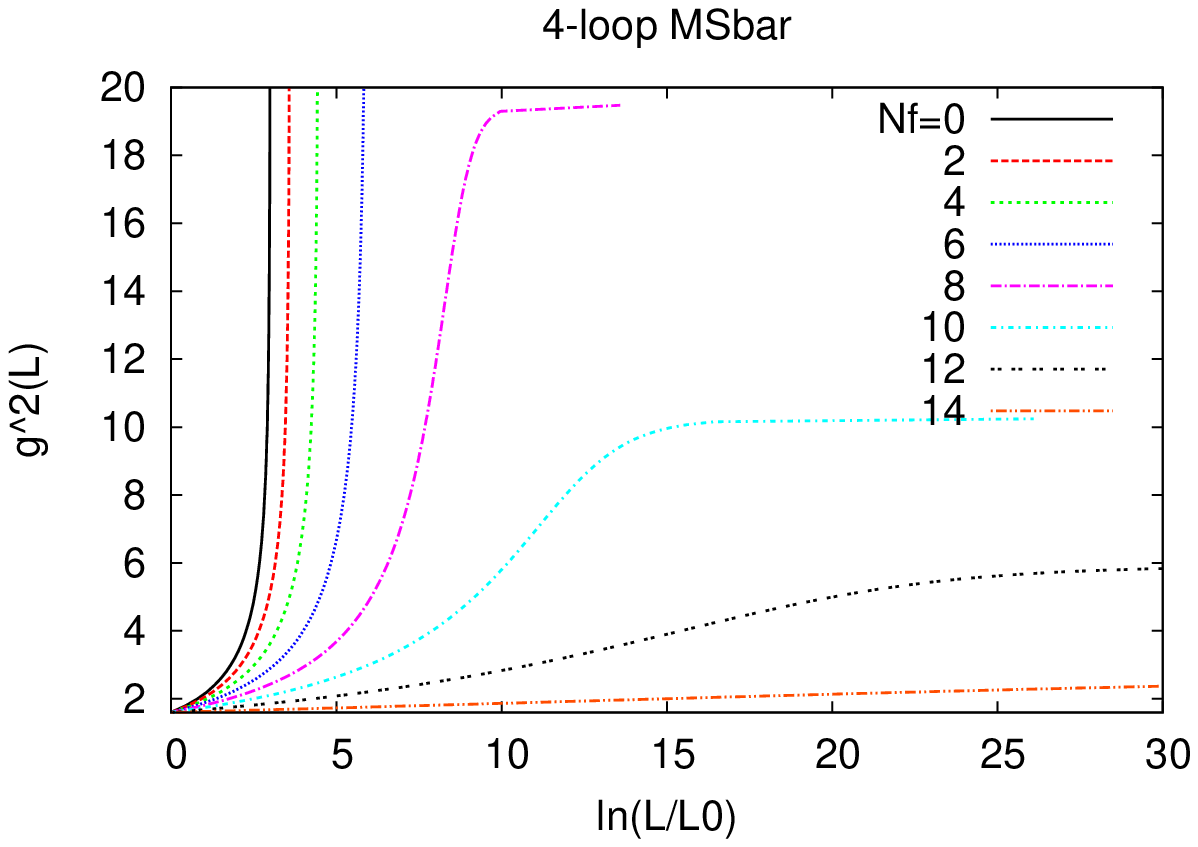}\\
\end{tabular}
\caption{$L$-dependence of $g^2(L)$ in different approximations or
 schemes.
 $N_f$ dependence is also shown.}
\label{fig:log_vs_g2}
\end{figure}
%%%%%%%%%%%%%%%%%%%%%%%
%%%%%%%%%%%%%%%%%%%%%%%
\begin{table}[tb]
\centering
\begin{tabular}{c|ccccccc}
 $N_f$ & 4 & 6 & 8 & 10 & 12 & 14 & 16\\
\hline
 2-loop universal &&&& 27.74 & 9.47 & 3.49 & 0.52\\
 \hline
 3-loop SF & 43.36 & 23.75 & 15.52 & 9.45 & 5.18 & 2.43 & 0.47\\
 \hline
 3-loop $\msbar$ && 159.92 & 18.40 & 9.60 & 5.46 & 2.70 & 0.50\\
 \hline
 4-loop $\msbar$ &&& 19.47 & 10.24 & 5.91 & 2.81 & 0.50\\
\end{tabular}
\caption{The IRFP from perturbative analysis.}
\label{tab:p-IRFP}
\end{table}
%%%%%%%%%%%%%%%%%%%%%%%
Fig.~\ref{fig:log_vs_g2} shows $L$-dependence of $g^2(L)$ in the four different schemes/approximations, i)-iv), with the initial condition $g^2(L_0)=1.6$.
The values of perturbative infrared fixed point (IRFP) are summarized in Tab.~\ref{tab:p-IRFP}.
As seen from Tab.~\ref{tab:p-IRFP}, once one goes beyond the two-loop approximation, the fixed point value of 12-flavor QCD is stable at around $g^{2}_{\rm IR}\sim$ 5 against the change of schemes/approximations.
It is interesting that this IRFP is completely consistent with that obtained in the non-pertrurbative calculation by~\cite{Appelquist:2007hu}.
Now looking at the perturbative IRFP at $N_{f}=10$, the similar stability is seen at $g^{2}\sim 10$.
Furthermore, according to an analysis based on the Shwinger-Dyson equation, there is an argument that chiral symmetry breaking occurs at around $g^{2}\sim 4\pi^{2}/(3\,C_{2}(R)) = \pi^2$~\cite{Appelquist:1988yc}.
In summary, the perturbative analysis suggests that 10-flavor QCD is the most attractive candidate for WTC.

\section{Simulation details}

We employ the \SF method~\cite{Luscher:1992an} to determine the running coupling constant.
Unimproved Wilson fermion action and the plaquette gauge action without any boundary counter terms are adopted to describe regularized dynamics of techniquarks and technigluons, respectively.
The parameter of the spatial boundary condition for fermions, $\theta$, is set to 0.
To determine the scale dependence, calculations on several different lattice sizes are required.
In this analysis, we report the results obtained with $(L/a)^{4}=4^{4}$, $6^{4}$, $8^{4}$ and $12^{4}$ lattices.
The calculation on $16^{4}$ lattice is in progress.
The bare gauge coupling $\beta=6/g^{2}_{0}$ is explored in the range of 4.4--24.0.
At around $\beta\sim$ 4.4, we encounter a bulk phase transition, where the plaquette value suddenly jumps to a smaller value.
Whenever this happens, we discard the configuration.
The numerical simulation is carried out on several architectures including GPGPU and PC cluster.
The standard HMC algorithm is used with some improvements in the solver part.
Since the Wilson type fermion explicitly violates chiral symmetry, the critical value of $\kappa$ has to be tuned to the massless limit.
We performed this tuning for every pair of $(\beta,\ L/a)$.
So far, we have accumulated 5,000 to 200,000 trajectories depending on $(\beta,\ L/a)$.

The procedure to determine the running is standard and consists of the following steps.\\[-5ex]
\begin{enumerate}
\setlength{\parskip}{0cm}
\setlength{\itemsep}{0cm}
\item Calculate the \SF scheme coupling at various $g^{2}_{0}$'s on several $L/a$'s, which is denoted by $g^{2}_{L/a}(g^{2}_{0})$.
\item Fit $g^{2}_{L/a}(g^{2}_{0})$ at each $L/a$ as a function of $g^{2}_{0}$ to obtain an interpolating formula $g^{2, \rm f it}_{L/a}(g^{2}_{0})$.
\item Set the input value $u_{0}$, which implicitly sets an initial length scale $L_{0}$. 
\item Choose one $L/a$ and determine $g^{2,*}_{0}$ satisfying $g^{2, \rm fit}_{L/a}(g^{2,*}_{0}) = u_{0}$.
\item Choose one $L'/a=sL/a$ (typically $s$ =2 or 3/2) and read the value of $g_{sL/a}^{2, \rm fit}(g^{2,*}_{0})$, which is stored in $\Sigma(sL_{0}/a,\,L_{0}/a)$.
\item Make a perturbative correction via
  $\Sigma(sL_{0}/a,\,L_{0}/a) = g_{sL/a}^{2, \rm fit}/[1+ \delta(a/sL_{0},\,a/L_{0})\times u]$ with the known coefficient $\delta$.
\item Repeat 4-6 with different $a$, {\it i.e.} for $L/a'$ and $L'/a'=sL/a'$.
\item Take the continuum limit by using $\Sigma(sL_{0}/a,\,L_{0}/a)$ and $\Sigma(sL_{0}/a',\,L_{0}/a')$ to obtain $\sigma(sL_{0})$.
\item Set $u_{1} = \sigma(sL_{0})$
\item Repeat 4-9 several times with the initial value $u_{1},\ u_{2},\ \cdots$. Then we obtain a series of
$\sigma(sL_0),\ \sigma(s^2 L_0),\ \sigma(s^3 L_0),\ \cdots$. This scale dependence describes the running of the coupling.
\end{enumerate}

\section{Results}

%%%%%%%%%%%%%%%%%%%%%%%
\begin{figure}[bt]
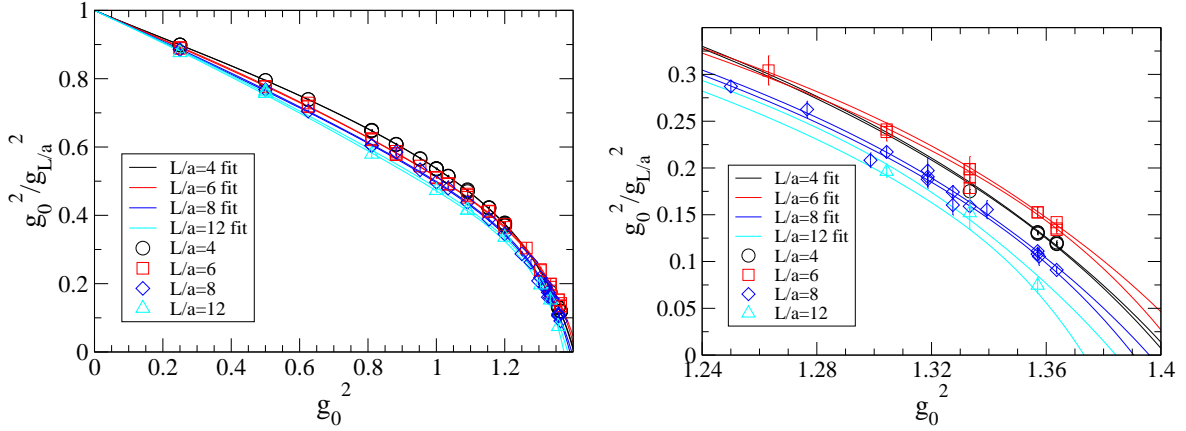

\centering
\begin{tabular}{cc}
\includegraphics*[width=0.5 \textwidth,clip=true]
{figs/fig_betavsg2_all.eps} &
\includegraphics*[width=0.5 \textwidth,clip=true]
{figs/fig_betavsg2_all_8.eps}
\end{tabular}
\caption{$g_{0}^{2}$-dependence of $g_{0}^{2}/g^2(L)$ for $L/a=4,\ 6,\ 8,\ 12$. The right panel is just an enlargement of the left.}
\label{fig:main_result_1}
\end{figure}
%%%%%%%%%%%%%%%%%%%%%%%
\begin{table}[htb]
\centering
\begin{tabular}{cc|ccccc}
$L/a$ & $N$ & $\chi^2$/dof & $a_{L/a,1}$ & $a_{L/a,2}$ & $a_{L/a,3}$ & $a_{L/a,4}$\\
\hline
  4 &  2  & 45.6(2.2) &  0.4903( 0.0006) & -0.4677( 0.0025)\\
  4 &  3  & 11.0(1.1) &  0.5003( 0.0034) & -0.3909( 0.0354) & -0.1038( 0.0435)\\
  4 &  4  & 2.2(0.5) &  0.5080( 0.0008) & -0.4746( 0.0094) &  0.1168( 0.0232) & -0.1486( 0.0148)\\
\hline
  6 &  2  & 19.5(1.6) &  0.4607( 0.0016) & -0.3826( 0.0038)\\
  6 &  3  & 4.3(0.8) &  0.4888( 0.0015) & -0.2662( 0.0085) & -0.1856( 0.0115)\\
  6 &  4  & 2.1(0.6) &  0.5028( 0.0026) & -0.4675( 0.0422) &  0.2486( 0.0894) & -0.2593( 0.0516)\\
\hline
  8 &  2  & 12.4(1.5) &  0.4822( 0.0018) & -0.4004( 0.0069)\\
  8 &  3  & 1.9(0.6) &  0.5088( 0.0021) & -0.2135( 0.0192) & -0.2668( 0.0239)\\
  8 &  4  & 0.7(0.4) &  0.5153( 0.0020) & -0.3610( 0.0515) &  0.0709( 0.1108) & -0.1972( 0.0605)\\
\hline
 12 &  2  & 9.7(1.9) &  0.4821( 0.0047) & -0.3870( 0.0162)\\
 12 &  3  & 2.7(1.0) &  0.5164( 0.0040) & -0.1716( 0.0316) & -0.3104( 0.0363)\\
 12 &  4  & 1.5(0.8) &  0.5269( 0.0033) & -0.3959( 0.0740) &  0.2611( 0.1728) & -0.3495( 0.1003)\\
\end{tabular}
\caption{The coefficients determined in the fit.}
\label{tab:fit_param1_pade_4}
\end{table}

%%%%%%%%%%%%%%%%%%%%%%%
Fig.~\ref{fig:main_result_1} essentially shows the \SF coupling obtained on the four lattices and their fit curves.
Notice that, instead of $g^{2}_{L/a}$, we deal with $1/g^{2}_{L/a}$ just for convenience.
As an interpolating function, we chose
\begin{eqnarray}
     \frac{g_0^2}{g_{L/a}^{2, \rm fit}(g^{2}_{0})}
 &=& \frac{1-a_{L/a,1}\,g_0^4}
          {1+p_{1,L/a}\times g_0^2+%\displaystyle
           \sum_{n=2}^N a_{L/a,n} \times g_0^{2\,n}
          },
 \label{eq:fitfunc}
\end{eqnarray}
where $p_{1,L/a}$ is the $L/a$-dependent, one-loop coefficient, the values of which for various $L/a$ are
\begin{eqnarray}
 p_{1,L/a} = \left\{
 \begin{array}{ll}
0.4225003137 & \mbox{ for } L/a=4\\
0.4477107831 & \mbox{ for } L/a=6\\
0.4624813408 & \mbox{ for } L/a=8\\
0.4756888260 & \mbox{ for } L/a=12 %\\
%0.4833079203 & \mbox{ for } L/a=16
 \end{array}
 \right..
\end{eqnarray}
We optimize the degree of polynomial in the denominator of (\ref{eq:fitfunc}), $N$, by monitoring $\chi^{2}$/dof.
The results of the fits are tabulated in Tabs.~\ref{tab:fit_param1_pade_4}.
From the $\chi^2$/dof values, $N$=4 is chosen in all the cases.

%%%%%%%%%%%%%%%%%%%%%%%
\begin{figure}[bt]
\centering
\begin{tabular}{c}
\includegraphics*[width=0.6 \textwidth,clip=true]
{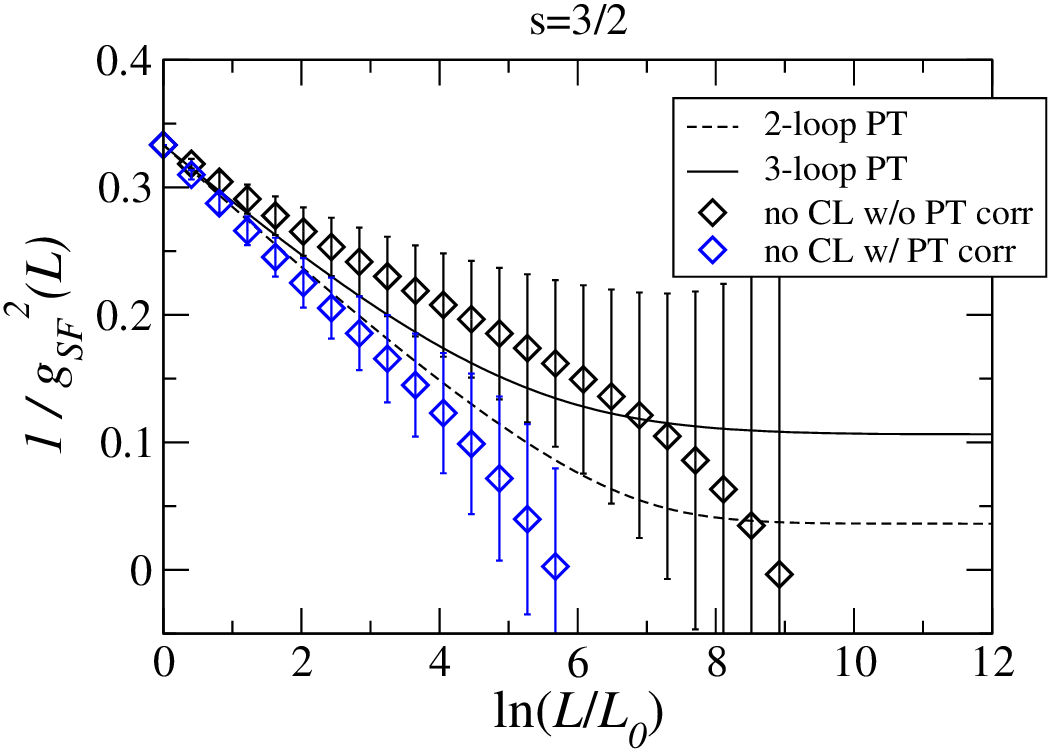}\\
\end{tabular}
\caption{Running of \SF coupling.}
\label{fig:main_result_2}
\end{figure}
%%%%%%%%%%%%%%%%%%%%%%%
Setting the initial value of the running $u_{0}$ to 3.0, we calculated the running through the procedure explained above.
Since we did not implement any $O(a)$ improvements, large scaling violation is expected.
This can be improved only by performing simulations at larger lattices.
We tried to evaluate the size of the scaling violation, but it fails because of the large statistical error.
Thus, in this report, instead of the continuum limit, the step scaling function constructed from the data on $L/a$=8 and 12 lattices is taken as the current estimate.
A subtlety occurs in the perturbative correction, {\it c.f.} the step 6 in the procedure.
While in the weak coupling region $\sigma(L)\le 2$ it works, in the other region the correction turns out to enhance the scaling violation.
This happens when the direction of the scaling violation at the one-loop level is opposite to that of the real violation.
Thus, we calculate the running coupling with and without the perturbative correction and interpret the difference as the potential systematic uncertainty associated with the scaling violation.
Fig.~\ref{fig:main_result_2} shows the result thus obtained.
The perturbative predictions with $\theta=\pi/5$ is also shown for a reference.
Although the statistical error is still large, an interesting observation is that the data without the perturbative correction almost goes along the 3-loop curves and at $g^{2}_{\rm SF}\sim 10$ turns to rapid increase.

\section{Summary and Outlook}

The running coupling constant of 10-flavor QCD is studied. 
The perturbative analysis suggests that this theory is extremely interesting.
The preliminary result obtained without continuum limit shows an interesting behavior.
In order to draw definite conclusions, we clearly need larger lattices and more statistics.
The calculation is in progress.

\vspace{3ex}\
Numerical simulations are performed on Hitachi SR11000
at High Energy Accelerator Research Organization (KEK) under a support of its Large Scale Simulation Program (No. 09-05), on GCOE (Quest for Fundamental Principles in the Universe) cluster system at Nagoya University and on the INSAM (Institute for Numerical Simulations and Applied Mathematics) GPU cluster at Hiroshima University.
This work is supported in part by the Grant-in-Aid for Scientific Research
of the Japanese Ministry of Education, Culture, Sports, Science and Technology
(Nos. 20105001, 20105002, 20105005, 21684013, 20540261 and 20740139), and
by US DOE grant \#DE-FG02-92ER40699.

\end{document}